\documentclass[aps,prl,twocolumn,superscriptaddress,nofootinbib]{revtex4-2}
\usepackage[utf8]{inputenc}
\usepackage[english]{babel}
\usepackage[expansion,protrusion]{microtype}
\usepackage{amsmath}
\usepackage{amssymb}
\usepackage{siunitx}
\usepackage{graphicx}
\usepackage[colorlinks=true,citecolor=blue,linkcolor=blue,urlcolor=blue]{hyperref}
\usepackage{booktabs}
\usepackage{eso-pic}

\DeclareMathOperator\real{Re}
\DeclareMathOperator\imag{Im}
\DeclareSIUnit{\muB}{\ensuremath{\mu_B}}

\begin{document}

\title{Engineering correlated Dirac fermions and flat bands on SiC\texorpdfstring{\\}{} with transition-metal adatom lattices}

\author{Henri Menke}
\thanks{These authors contributed equally}
\affiliation{Department of Physics, Friedrich-Alexander-Universit\"at Erlangen-N\"urnberg, 91058 Erlangen, Germany}
\affiliation{Max Planck Institute for Solid State Physics, Heisenbergstr.~1, 70569 Stuttgart, Germany}

\author{Niklas Enderlein}
\thanks{These authors contributed equally}
\affiliation{Department of Physics, Friedrich-Alexander-Universit\"at Erlangen-N\"urnberg, 91058 Erlangen, Germany}

\author{Roland Gillen}
\affiliation{Department of Physics, Friedrich-Alexander-Universit\"at Erlangen-N\"urnberg, 91058 Erlangen, Germany}
\affiliation{College of Engineering, Swansea University, Swansea SA1 8EN, United Kingdom}

\author{Yi-Ting Tseng}
\affiliation{Department of Physics, Friedrich-Alexander-Universit\"at Erlangen-N\"urnberg, 91058 Erlangen, Germany}

\author{Michel Bockstedte}
\affiliation{Institute for Theoretical Physics, Johannes Kepler University Linz, Altenbergerstr.~69, 4040 Linz, Austria}

\author{Janina Maultzsch}
\affiliation{Department of Physics, Friedrich-Alexander-Universit\"at Erlangen-N\"urnberg, 91058 Erlangen, Germany}

\author{Giorgio Sangiovanni}
\affiliation{Institut f\"ur Theoretische Physik und Astrophysik and W\"urzburg-Dresden Cluster of Excellence ct.qmat, Universit\"at W\"urzburg, 97074 W\"urzburg, Germany}

\author{Philipp Hansmann}
\email{philipp.hansmann@fau.de}
\affiliation{Department of Physics, Friedrich-Alexander-Universit\"at Erlangen-N\"urnberg, 91058 Erlangen, Germany}

\date{\today}

\begin{abstract}
We propose three transition-metal adatom systems on SiC surfaces as a versatile platform to realize massless Dirac fermions and flat bands with strong electronic correlations.
Using density functional theory combined with the constrained random phase approximation and dynamical mean-field theory, we investigate the electronic properties of~Ti, V, and~Cr adatoms.
The triangular surface lattices exhibit narrow bandwidths and effective two-band Hubbard models near the Fermi level, originating from partially filled adatom $d$-orbitals.
For the undoped systems our calculations reveal two distinct Mott insulating ground states. While the V lattice is a paramagnetic textbook case with large local moments, the Cr lattice, in contrast, is on the edge of a phase transition towards a flat-band Fermi liquid. The Ti lattice realizes a heavy Dirac semimetal at zero doping.
\end{abstract}

\maketitle

The interplay of Dirac fermions, flat bands, and strong correlations is central to condensed-matter physics, driving novel quantum phases and exotic material properties.
For transport applications, these features must lie sufficiently close to the Fermi level ($E_\text{F}$).
Twisted bilayer graphene and kagome materials~\cite{YCao2018,Andrei2021,TBG_Bernevig_Yazdani,lisiNatPhys,KangNatMat2020,YinReviewKagome,Cr135_Nature,GiorgioNV} have shown significant progress; however, both face limitations in scalability or tunability. The former requires complex stacking procedures not suited for large-area applications, the latter allows only partial van Hove singularity tuning near $E_\text{F}$ via strain or doping.
Thus, material platforms for studying correlated relativistic fermions and flat-band-induced many-body instabilities remain scarce, highlighting the need for new systems that naturally host these phenomena without such constraints.

Adatom lattices on semiconductor substrates can host isolated flat bands.
Group-IV adatoms (Sn, Pb, C) on Si(111) and Ge(111) at 1/3 monolayer (ML) coverage~\cite{ganz1991submonolayer,carpinelli1996direct,carpinelli1997surface,profeta2005novel,profeta2007triangular,schuwalow2010realistic,chaput2011giant,li2011geometrical,hansmann2013jpcm,hansmann2013prl,hansmann2016,gangli2013} realize extended single-band Hubbard models on triangular lattices. These systems, when doped, exhibit phenomena like chiral superconductivity as theoretically predicted for these specific compounds~\cite{cao2018} and experimentally observed~\cite{weitering2020,weitering2023} in Sn:Si(111).
Other adatom systems, such as hexagonal bilayers and In-monolayers on 4H-SiC(0001), show multi-orbital topological band structures~\cite{reis2017bismuthene,bauernfeind2021design}. However, strong spin-orbit coupling (SOC) and substrate-induced symmetry breaking open bulk gaps that reduce their Mott-like character.
Thin-layer transition-metal (TM) dichalcogenides are another candidate, but chalcogen $p$-hybridization and SOC suppress correlated/topological features.~\cite{Kennes2021,Tang2020}.

\begin{figure*}[t]
\includegraphics[width=\textwidth]{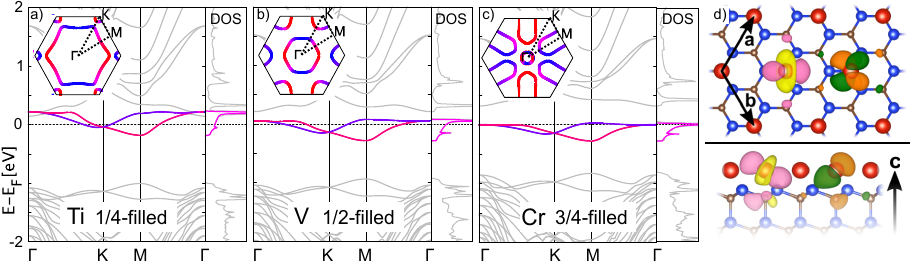}
\caption{a)-c): DFT bands (grey) for Ti, V, Cr adatom lattices overlaid with downfolded two-band Wannier model. The generally mixed orbital character is color-coded: blue (pure $z^2$-like) / red (pure ${x^2-y^2}$-like); same color scheme is used for the FS (insets). The DOS has identical contributions from both (locally degenerate) Wannier orbitals. d): $\sqrt{3} \times \sqrt{3}$ surface unit cell at 1/3 ML adatom coverage; red spheres mark T$_4$ adatom sites with the two localized Wannier orbitals shown.
}
\label{fig:dft}
\end{figure*}

Here, we propose a new adatom-based platform using 3$d$ TM adatoms on SiC.
Specifically, Ti, V, and Cr adatoms on a 3C-SiC(111) surface (or related (0001) 4H/6H-SiC) emerge as versatile systems to study triangular-lattice magnetic moments, Dirac fermions, and flat bands in a strongly correlated regime.
SiC offers a robust, wide-gap substrate with strong bonds that suppress adatom diffusion, thus promoting stability~\cite{reis2017bismuthene,bauernfeind2021design,di2019towards,gehrig2025bismuthene,glass2015triangular}.
Using 3$d$ TMs leverages inherent electronic correlations, including multi-orbital degrees of freedom and Hund's physics~\cite{demedici2011janus, georges_kotliar2024}.
At 1/3~ML coverage, these TMs form stable triangular lattices with $\sqrt{3}\times\sqrt{3}$ geometry, hosting two narrow surface bands with Dirac cones near $E_\text{F}$.
This contrasts with systems like graphene or 3D topological insulators, where Dirac cones lack the strong correlations needed for emergent many-body physics~\cite{Hofmann2014}.
Our results reveal a filling-dependent trend: Ti realizes a correlated Dirac Fermi liquid, V lies deep in a paramagnetic Mott phase with large local moments, while Cr is a Mott insulator on the verge of a phase transition: a small amount ($<10$\%) of hole- or electron doping can drive the systems into Fermi liquid metallic phases with flat-band or heavy Dirac quasiparticles. This control makes them an ideal platform to study the interplay of topology and correlations via doping-induced metallization.

We combine density functional theory~(DFT), constrained random phase approximation~(cRPA), and fully charge self-consistent (CSC) dynamical mean-field theory~(DMFT).
As our proposed TM adatom lattices have not yet been realized experimentally, we assess their structural stability (beyond simple DFT relaxation) via phonon spectra and molecular dynamics (see~\cite{Supplemental} for details). Both methods confirm dynamical stability, supporting experimental feasibility.
The screened Coulomb interaction parameters were computed with cRPA~\cite{aryasetiawan2004frequency,vaugier2012hubbard} to construct a Hubbard-Kanamori Hamiltonian for DMFT:

\begin{equation}
\begin{aligned}
\hat{H}_{\mathrm{int}} = &\ U \sum_m \hat{n}_{m\uparrow} \hat{n}_{m\downarrow}
+ U' \sum_{m\neq m'} \hat{n}_{m\uparrow} \hat{n}_{m'\downarrow} \\
&+ (U' - J_{\mathrm{H}}) \sum_{m < m', \sigma} \hat{n}_{m\sigma} \hat{n}_{m'\sigma} \\
&- J_{\mathrm{H}} \sum_{m \neq m'} \hat{c}^\dagger_{m\uparrow} \hat{c}_{m\downarrow}
\hat{c}^\dagger_{m'\downarrow} \hat{c}_{m'\uparrow} \\
&+ J_{\mathrm{H}} \sum_{m \neq m'} \hat{c}^\dagger_{m\uparrow} \hat{c}^\dagger_{m\downarrow}
\hat{c}_{m'\downarrow} \hat{c}_{m'\uparrow}
\end{aligned}
\end{equation}
with intra-/inter-orbital interaction $U/U'$ and Hund's coupling $J_{\mathrm{H}}$.
\begin{figure*}[t]
\includegraphics[width=2\columnwidth]{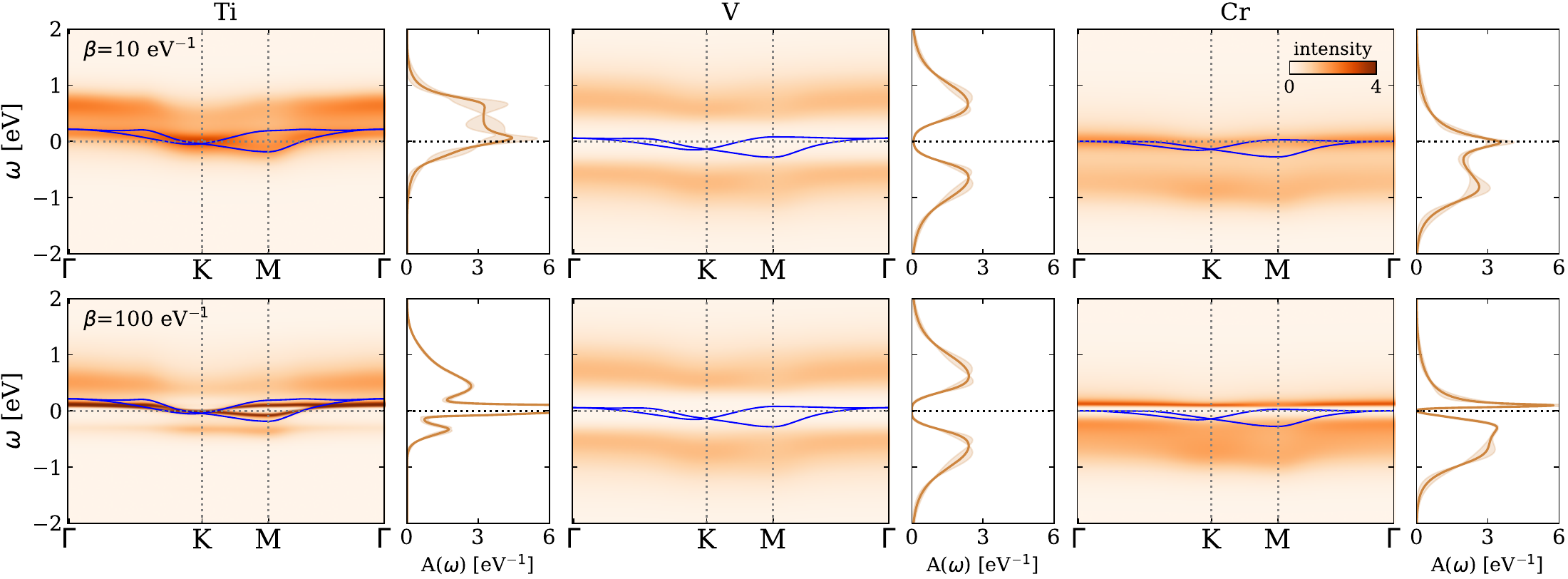}
\caption{\label{fig:dmft_akw} DMFT spectral functions for the three adatom species Ti, V, Cr (columns) and two inverse temperatures $\beta = \SI{10}{\per\eV}$ and $\SI{100}{\per\eV}$ (rows). Momentum-resolved spectral functions are shown via color scale; blue lines denote the (uncorrelated) Wannier bands. To the right of each panel is the $k$-integrated spectral function (DOS), with lines indicating the average and the transparent envelope denoting analytic continuation error bars.}
\end{figure*}
We then performed CSC DFT+DMFT~\cite{Anisimov1997, Lichtenstein1998, dmft3} using the \textsc{solid\_dmft} package, interfacing Quantum ESPRESSO (DFT) and the TRIQS library (DMFT)~\cite{Merkel2022, Beck2022}. SOC was neglected. Further details can be found in the supplemental material~\cite{Supplemental}.

Fig.~\ref{fig:dft} a)-c) shows DFT bands along the high-symmetry path ($\Gamma$-$K$-$M$-$\Gamma$) in the hexagonal Brillouin zone, with Fermi surfaces (FS; insets) and density of states (DOS). All systems exhibit similar band structures, featuring two well-isolated, narrow bands near $E_\text{F}$. The van Hove singularities in the DOS are typical of triangular lattices with both nearest and next-nearest neighbor hopping~\cite{hansmann2013jpcm}.

The two associated Wannier orbitals, mainly from the adatom $3d$-shell, belong to a degenerate ``$E$'' representation of the trigonal point group at the T$_4$ adatom site (see Fig.~\ref{fig:dft}~d) and section II.\ in~\cite{Supplemental}). These orbitals, resembling tilted 3$z^2-\mathbf{r}^2$ and $x^2-y^2$ types, contain subsurface contributions but remain localized, with spreads of \SI{3.7}{\angstrom}~(Ti), \SI{2.7}{\angstrom}~(V), and \SI{2.3}{\angstrom}~(Cr), reflecting the narrow bandwidths of \SI{0.40}{\eV}~(Ti), \SI{0.36}{\eV}~(V), and \SI{0.30}{\eV}~(Cr) which indicate potential for strong correlations.

The triangular symmetry of the adatom lattice gives rise to notable dispersion relations, including Dirac cones near the $K$ point and a very flat band around $M$-$\Gamma$, contributing to a sharp DOS peak.
In the absence of SOC the Dirac points are protected by the out-of-plane mirror symmetry~\cite{Eck2022} present along the connectors of nearest-neighbor T$_4$ adsorption positions (see supplemental material~\cite{Supplemental}).

In the real-space Wannier basis, the adatom bands strongly mix, as seen in the color-coded orbital character of bands and FS in Fig.~\ref{fig:dft} a)-c). The orbital composition, however, is not unique, depending on the chosen local reference (i.e., rotations within the irreducible representation subspace).
While dispersion relations are qualitatively similar, the effective band filling varies, leading to distinct FS. The $1/4$-filled Ti system (nominally in a 3d$^1$ configuration) has a large $\Gamma$-centered hexagonal sheet and six small circular pockets near the $K$ points. For half-filled V (3d$^2$), the FS retains this symmetry but shows larger $K$-point pockets and a smaller $\Gamma$-centered hexagon. At $3/4$ filling (Cr; 3d$^3$), a Lifshitz transition now produces a FS of six U-shaped $M$-pockets and small circular $\Gamma$-pocket.
These FS differences underline the complex interplay between adatom species and electronic structure, with filling control via adatom choice as key tuning parameter. It enables a coarse setting of the filling that can be refined through additional chemical doping (e.g., boron hole doping), offering access to a wider filling range than doping alone, which becomes increasingly impractical beyond 10\% due to disorder and other limitations~\cite{weitering2020}.

\begin{table}[b]
\centering
\caption{Screened and unscreened Hubbard-Kanamori parameters (intra-/inter-orbital $U$/$U'$ and Hund's coupling $J_{\mathrm{H}}$) from cRPA and averaged nearest-neighbor interaction $U^{nn}$.}
\begin{tabular*}{\linewidth}{@{\extracolsep{\fill}}c l S[table-format=2.2] S[table-format=2.2] S[table-format=1.2] S[table-format=1.2] @{}}
\toprule
adatom & (in unit \si{\eV}) & {$U$} & {$U^{\prime}$} & {$J_{\mathrm{H}}$} & {$\langle U^{nn}\rangle$} \\
\midrule
       & screened     & 0.86  & 0.57  & 0.14   & 0.44 \\
Ti     & unscreened   & 7.22  & 6.72  & 0.25  &  2.72 \\
       & ratio        & 0.12  & 0.09  & 0.57  & 0.16 \\
\hline
       & screened     & 1.18  & 0.69  & 0.24   & 0.43 \\
V   & unscreened   & 10.90  & 10.13  & 0.38  & 2.71  \\
       & ratio        & 0.11  & 0.07  & 0.64  & 0.16 \\
\hline
       & screened     & 1.40  & 0.80  & 0.30   & 0.47 \\
Cr & unscreened   & 12.85  & 11.92  & 0.46  & 2.70 \\
       & ratio        & 0.11  & 0.07  & 0.65  & 0.17 \\
\bottomrule
\end{tabular*}
\label{tab:cRPA}
\end{table}

Next, we computed effective screened interaction parameters for the correlated subspace spanned by the two Wannier orbitals using cRPA. Tab.~\ref{tab:cRPA} summarizes both screened and bare interaction parameters.
For all systems, the effective onsite interaction~$U$ is large relative to the narrow bandwidths, indicating strong correlation effects. The interactions follow the expected trend: $U$ reaches \SI{1.40}{\eV} for the most localized orbitals in Cr, with smaller values of $U = \SI{1.18}{\eV}$ for V and $U = \SI{0.86}{\eV}$ for Ti.
Additionally, we find sizable nearest-neighbor interactions $\langle U^{nn}\rangle$, comparable to prior cRPA results for $\alpha$-phases on Si(111)~\cite{hansmann2013jpcm}. Overall, the interaction parameters are similar to those of C:Si(111), where nearest-neighbor and longer-range interactions were, however, shown to have negligible effects beyond mean-field levels~\cite{hansmann2013prl}. Thus, we considered only local interactions in DMFT.

In Fig.~\ref{fig:dmft_akw}, we show momentum-resolved and local spectral functions from DMFT at inverse temperatures $\beta = \SI{10}{\per\eV}$ and \SI{100}{\per\eV} along the high-symmetry path from Fig.~\ref{fig:dft}, with uncorrelated DFT bands overlaid in blue for reference.
While the DFT bands appear similar across the three systems, the correlated spectral functions reveal substantial differences, underlining the role of filling.

\begin{figure}[t]
\includegraphics[width=\columnwidth]{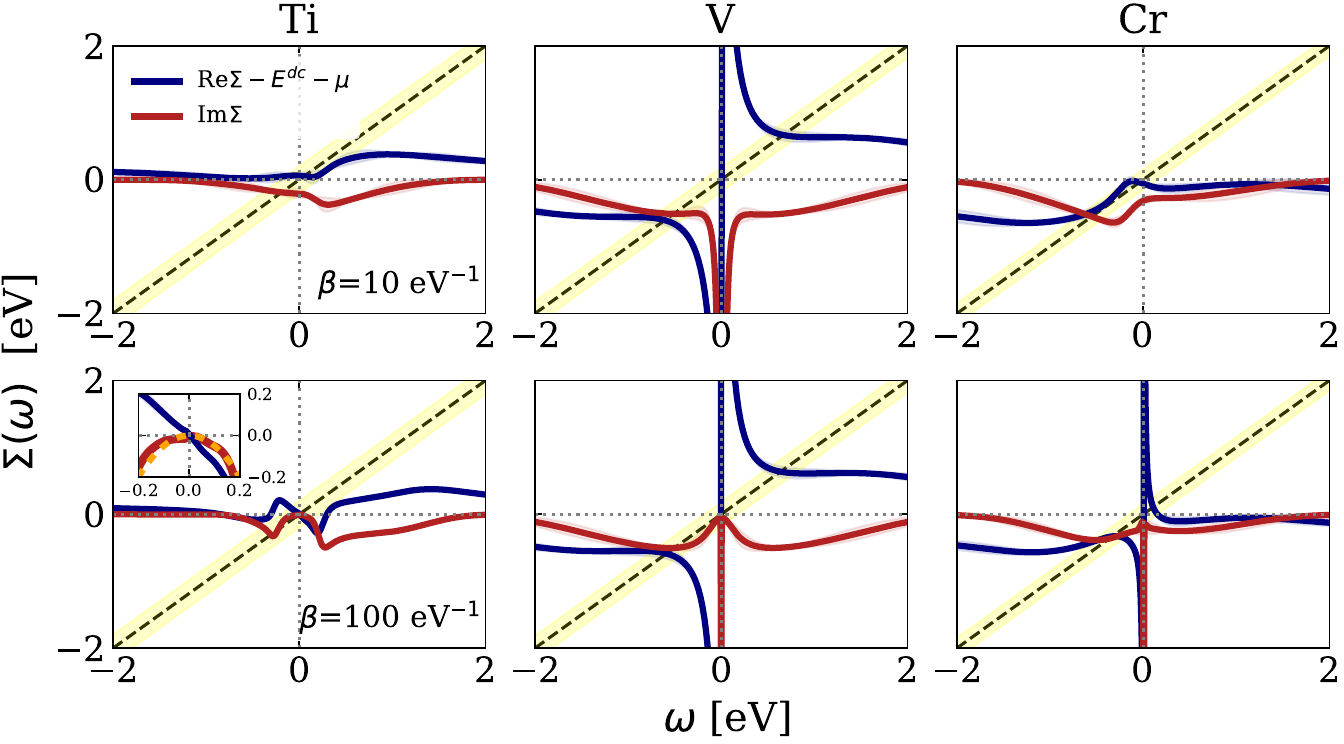}
\caption{\label{fig:dmft_sigma} DMFT self-energies (transparent envelopes represent analytic continuation error bars) corresponding to the spectral functions in Fig.~\ref{fig:dmft_akw}. Dashed line: angle bisector; yellow shaded area: width of non-interacting bands.
}
\end{figure}

For V, we observe a well-defined Mott insulating state at both temperatures, with distinct Hubbard bands and a \SI{1}{\eV} gap.
This Mott gap is a direct consequence of the the pole of the real part of the self-energy $\real{
\Sigma(\omega = 0)}$ (blue lines in central panels of Fig.~\ref{fig:dmft_sigma}). From the intersection of $\real{
\Sigma(\omega)}$ with the yellow shaded area (indicating the bandwidth of the non-interacting dispersion) around the angle bisector, we can deduce the position of the Hubbard bands.
We underline, that they emerge from the \textit{dynamical} nature of the DMFT self-energy which preserves a paramagnetic ground state. While the gap can also be captured using DFT+U, this requires a \textit{static} self-energy that breaks orbital- or spin symmetry. In the supplemental material we provide extensive DFT+U results on symmetry broken (i.e.\ magnetically ordered) ground states~\cite{Supplemental}. In this \emph{static} mean-field approach, we were able to stabilize various ordering patterns with comparable total energies which can be seen as indication to strong magnetic fluctuations in the real system.

Further, one might wonder if the Mott gap stems from an effective lowering of the lattice symmetries protecting the Dirac cone. This is not the case, as neither lattice nor time-reversal symmetries are broken upon entering the Mott phase, which is in line with recent studies on the zeros of the Green's function showing the existence of Dirac zeros in the gapped phase~\cite{Wagner2023, Blason2023,Setty2024,ivarAnom}.
To characterize the Mott insulating ground state of V further, we computed the dynamic impurity susceptibility~\cite{Hansmann2010, Toschi2012}:
\begin{equation}
\label{eq:chi}
    \chi_\text{loc}(\tau)
    \equiv g^2  \sum_{i,j} \langle \mathcal{T}_\tau
        S_z^i(\tau)\ S_z^{j}(0) \rangle,
\end{equation}
where $S_z^i(\tau)=1/2\big[ n^i_\uparrow (\tau) - n^i_\downarrow (\tau)\big]$ and $g$ is the gyromagnetic factor. The effective local moment is obtained as $m^\text{V}_\text{loc}\simeq \sqrt{3 \chi(\omega=0)}=\SI{2.81}{\muB}$, consistent with an ${S=1}$ localized moment (reflected in near-constant $\chi_\text{loc}(\tau)$).

\begin{figure}[t]
\includegraphics[width=0.95\columnwidth]{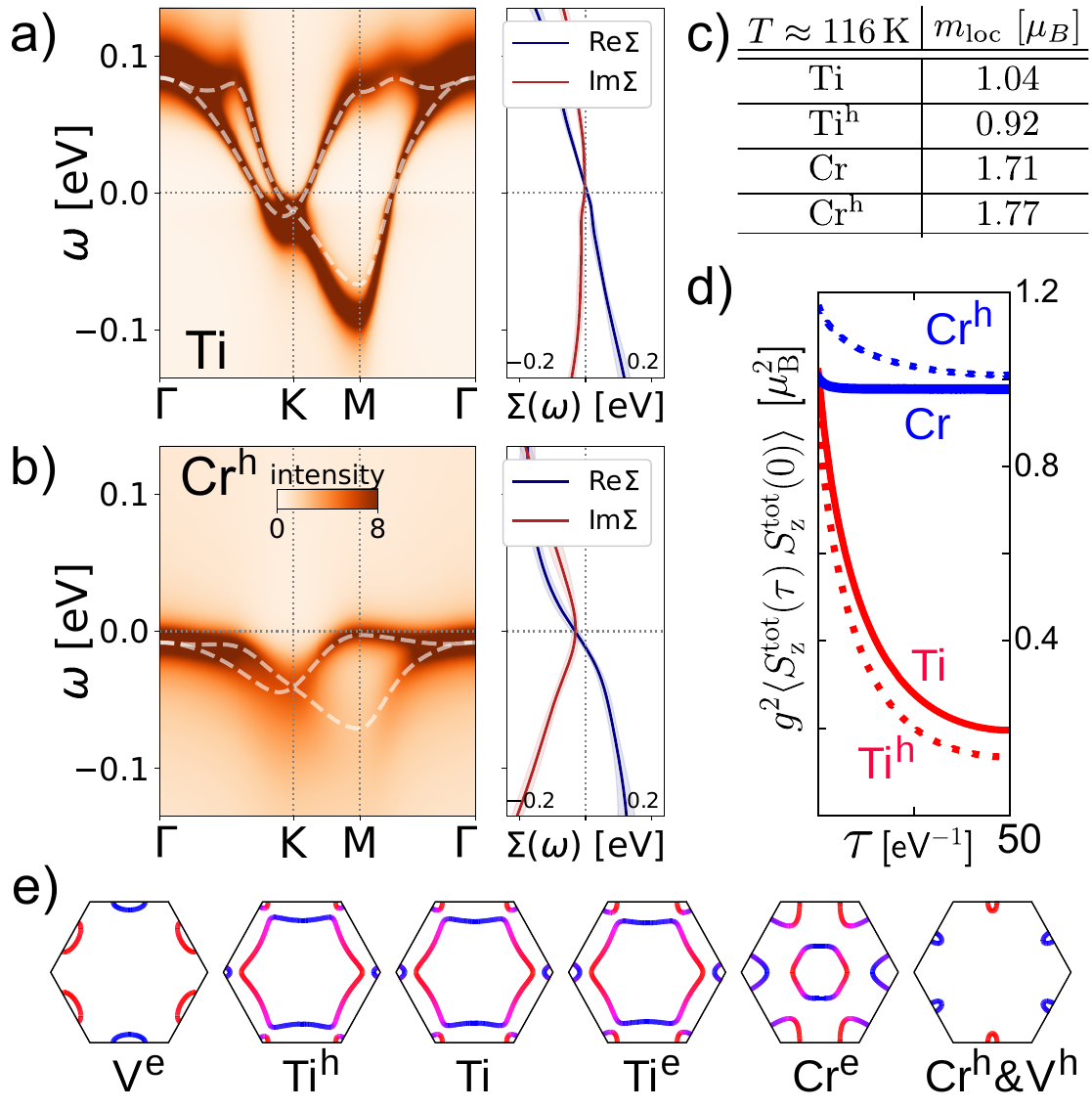}
\caption{\label{fig:dmft_akw_zoom} Renormalized DMFT quasiparticle bands of undoped Ti a) and hole-doped Cr$^h$ b) for $\beta = \SI{100}{\per\eV}$. Dirac cones at $K$ and the flat band around $M$-$\Gamma$ for Cr$^h$ are inherited from the non-interacting dispersion (whose renormalized version is indicated by white dashed lines). Side panels display the corresponding self-energies. c) Table of local magnetic moments for hole- and undoped systems. d) Spin-spin correlation functions: The operator $S^\text{tot}_z$ denotes the $z$-component of the spin summed over both orbitals. e) Renormalized FS of all metallic systems ordered according to their effective $E_\text{F}$.}
\end{figure}

Turning to Ti and Cr, in contrast, we find ``bad metal'' behavior at $\beta=\SI{10}{\per\eV}$, with finite, yet incoherent spectral weight near $E_\text{F}$ (see Fig.~\ref{fig:dmft_akw}) and high quasiparticle scattering rates ($\propto \imag\Sigma(\omega=0)$, see red lines in the upper row of Fig.~\ref{fig:dmft_sigma}).

At lower temperatures, Ti displays clean Fermi liquid behavior (i.e., $\real(\Sigma(\omega\approx 0))\propto \omega$; $\imag(\Sigma(\omega\approx 0)\propto \omega^2$) — see orange dashed parabola guiding the eye in the inset of Fig.~\ref{fig:dmft_sigma}, lower left) with a sharp quasiparticle peak at $E_F$. In contrast, Cr develops typical (narrow) Mott poles in $\real\Sigma(\omega=0)$ (Fig.~\ref{fig:dmft_sigma}, lower right) and a correspondingly small spectral gap (Fig.~\ref{fig:dmft_akw}, lower right), with sizable local magnetic moments of about $\SI{1.7}{\muB}$. The presence of significant spectral weight just above $E_F$ indicates that Cr lies close to an insulator-to-metal transition; indeed, both Cr and V can be driven into metallic phases by doping.

In subplots a)-b) of Fig.~\ref{fig:dmft_akw_zoom} we show sharp quasiparticle bands around $E_\text{F}$ of the undoped Ti and hole-doped Cr$^h$ lattices.
From the corresponding self-energies low scattering rates (small $\imag\Sigma(\omega=0)$) and strong mass renormalization $m^*/m=Z^{-1}$ with  $Z=(1-\partial_\omega\real\Sigma(\omega=0))^{-1}$ can be extracted with $m^*/m=2.7$ for Ti and $m^*/m=4.5$ for Cr$^h$.
While Ti displays quasiparticle Dirac cones in the immediate vicinity of $E_\text{F}$, Cr$^h$ features an extremely flat band, which, together with the correlation-induced mass renormalization, leads to an exceptionally high quasiparticle mass.
Note that - in contrast to the atomic limit - the quasiparticle dispersions on the triangular lattice in the metallic Ti and doped Cr systems break particle-hole symmetry, resulting in distinct magnetic responses of these strongly correlated metals.

In Fig.~\ref{fig:dmft_akw_zoom} c) we list local magnetic moments for hole- and undoped Ti and Cr, extracted from their dynamic spin susceptibility shown in panel d) as a function of imaginary time $\tau$.
This quantity characterizes the interplay of localized moments and renormalized quasielectrons on different time-scales (see~\cite{Hansmann2010, Toschi2012, Watzenboeck2020}).
The instantaneous response at $\tau=0$ reflects the bare local moment and the increase (decrease) of that response in Cr$^h$ (Ti$^h$) is intuitive: Cr$^h$ is driven towards the $S=1$ configuration we found for V while Ti$^h$ converges towards the empty $S=0$ limit.
We also observe a striking difference in the $\tau$-dependence, indicating how the bare moment is screened by the Fermi liquid on different time-scales: faster decay for Ti implies more efficient screening, likely due to its larger quasiparticle bandwidth (see panel a) compared to the flatter dispersion of Cr$^h$.

Fig.~\ref{fig:dmft_akw_zoom} e) shows the resulting quasielectron FS for all doped systems including their color-coded orbital character.
The distinct FS topologies can be characterized by counting the number of Dirac quasielectrons at $E_\text{F}$: Electron-doped Cr$^e$ and the Ti cases can be assigned a Berry phase of $\pi$ per spin and per valley, while for V and Cr$^h$ the FS encircles no Dirac point~\cite{haldane2004berry}.
Detailed DMFT spectra, magnetic responses for all, and FS properties of all metallic systems
can be found in the supplemental material~\cite{Supplemental}.

In conclusion, we have examined the electronic properties of Ti, V, and Cr adatom lattices on a 3C-SiC(111) surface through DFT, cRPA, and CSC DMFT. The charge feedback in the CSC had minimal impact, indicating that the relevant low-energy structure and many-body correlations are highly localized on the adatoms. These systems thereby realize clean, two-band Hubbard models — a rare and valuable feature.
We find V to realize a robust Mott-insulating state, suggesting possible spin-glass behavior on the triangular lattice. In contrast, Cr lies close to the correlated metallic regime and, upon slight hole doping, exhibits quasiparticle bands with pronounced flat-band features. Finally, the Ti system already forms a strongly correlated Fermi liquid in the undoped state, with heavy quasiparticle bands hosting Dirac cones near $E_\text{F}$.

For V, the particularly large local moments are notable as magnetic fluctuations were suspected to play a role in other (i.e.\ group IV) adatom systems~\cite{gangli2013} but have not yet been observed experimentally. Future research could explore heavier adatoms, where SOC may provide further degrees of freedom for tailoring correlated electronic structures. Altogether, our work unveils a remarkable versatility and tunability of two-dimensional TM adatom lattices on SiC which we propose as new platforms for exploring quantum phases of correlated electrons, including topological superconductivity.

\begin{acknowledgments}
We thank Alexander Hampel, Sophie Beck, Max Xylander, Federico Mazzola, Jonas Erhardt, Ralph Claessen, Carmine Ortix and Sabine Maier, for fruitful discussions. The authors gratefully acknowledge the scientific support and HPC resources provided by the Erlangen National High Performance Computing Center (NHR@FAU) of the Friedrich-Alexander-Universität Erlangen-Nürnberg (FAU) under the NHR project b181dc. NHR@FAU hardware is partially funded by the German Research Foundation (DFG) – 440719683. R.G. acknowledges the support of the Supercomputing Wales project, which is part-funded by the European Regional Development Fund (ERDF) via the Welsh Government. N.E. and J.M. acknowledge partial funding by the DFG Research Training Group GRK2861 – 491865171. Y.T. acknowledges financial support of the DFG project HA7277/3-1. M.B. received financial support from the Austrian Science Fund (FWF):I5195-N and German Research Foundation (DFG, QuCoLiMa, SFB/TRR 306, Project No. 429529648). G.S. acknowledges financial support from the DFG through the Wuerzburg-Dresden Cluster of Excellence on Complexity and Topology in Quantum Matter ct.qmat (EXC 2147, Project-id 390858490).
\end{acknowledgments}

\noindent\emph{Data availability} The data is available from the authors upon request. The codes for the DFT and DMFT calculations are available open source, the code for the cRPA calculations is available under commercial license.


%

\pagestyle{empty}
\makeatletter
\@for\next:=1,2,3,4,5,6,7,8,9,10,11,12,13,14,15\do{%
  \clearpage\leavevmode
  \edef\next{\noexpand\AddToShipoutPictureBG*{\noexpand\includegraphics[page=\next]{supplemental.pdf}}}%
  \next
}%

\end{document}